\documentclass[journal]{IEEEtran}
\usepackage{amssymb}
\usepackage[cmex10]{amsmath}
\usepackage{cite}
\usepackage{graphicx}
\usepackage[normalem]{ulem}

\usepackage{epstopdf}
\begin{document}
\title{Comprehensive Characterization of InGaAs/InP Avalanche Photodiodes at 1550\,nm with an Active Quenching ASIC}
\author{Jun~Zhang, Rob~Thew, Jean-Daniel~Gautier, Nicolas~Gisin, and Hugo~Zbinden
\thanks{Manuscript received. This work was supported by the Swiss NCCR Quantum Photonics and the Swiss CTI.}
\thanks{J.~Zhang, R.~Thew, J.-D.~Gautier, N.~Gisin, and H.~Zbinden are with the Group of Applied Physics, University of Geneva,
1211 Geneva 4, Switzerland e-mail: (Jun.Zhang@unige.ch).}}

\markboth{IEEE JOURNAL OF QUANTUM ELECTRONICS,~Vol.~, No.~}%
{Shell \MakeLowercase{\textit{et al.}}: Comprehensive Characterization of InGaAs/InP Avalanche Photodiodes at 1550~nm with an Active Quenching ASIC}

\maketitle

\begin{abstract}
We present an active quenching application specific integrated
circuit (ASIC), for use in conjunction with InGaAs/InP avalanche
photodiodes (APDs), for 1550\,nm single-photon detection. To
evaluate its performance, we first compare its operation with that
of standard quenching electronics. We then test 4 InGaAs/InP APDs
using the ASIC, operating both in the free-running and gated modes,
to study more general behavior. We investigate not only the standard
parameters under different working conditions but also parameters
such as charge persistence and quenching time. We also use the
multiple trapping model to account for the afterpulsing behavior in
the gated mode, and further propose a model to take account of the
afterpulsing effects in the free-running mode. Our results clearly
indicate that the performance of APDs with an on-chip quenching
circuit significantly surpasses the conventional quenching
electronics, and makes them suitable for
 practical applications, e.g., quantum cryptography.
\end{abstract}


\begin{IEEEkeywords}
Avalanche photodiodes (APDs), SPAD, single-photon detection, telecom wavelengths, ASIC, quantum cryptography.
\end{IEEEkeywords}

\IEEEpeerreviewmaketitle

\section{Introduction}

Single-photon detectors are the key components in numerous
photonics-related applications such as quantum
cryptography~\cite{GRTZ02}, optical time domain
reflectometry~\cite{OTDR,WSG04} and integrated circuit testing~\cite{IC}.
We can classify single-photon detection into four classes:
photomultiplier tubes~\cite{PMT}; semiconductor
APDs~\cite{CGLSZ96,Cam04}; superconducting detectors~\cite{SSPD};
and novel proposals such as using a single-electron transistor
consisting of a semiconductor quantum dot~\cite{SET}. In the
telecommunication regime (1550\,nm), InGaAs/InP APDs are currently
the best choice for practical applications such as quantum
cryptography~\cite{GRTZ02} due to their favorable characteristics
such as cost, size and robust operation with only thermo-electric
cooling required.

To detect single photons, APDs must work in the so-called Geiger
mode in which an inverse bias voltage ($V_{bv}$), exceeding the
breakdown voltage ($V_{br}$), is applied, such that even a single photoexcited
carrier (electron-hole pair) can create a persistent avalanche and a
subsequent macroscopic current pulse due to the process of impact
ionization. After the avalanche, a passive or active quenching
circuit~\cite{CGLSZ96}, is used to reduce $V_{bv}$ down to below
$V_{br}$, output a synchronized pulse and reset the APD for
detecting the next photon.

InGaAs/InP APDs are currently  fabricated with separate absorption,
charge and multiplication layers~\cite{Cam04} to ensure the
lattice matching and preserve a low electric field in the InGaAs
absorption layer with a narrower bandgap ($E_g=0.75$\,eV for
In$_{0.53}$Ga$_{0.47}$As), minimizing the induced leakage
currents, while a high electric field in the InP multiplication layer,
enhancing the impact ionization effect. The middle charge layer
can efficiently control the electric field profiles of the
absorption and multiplication layers. The parameters of APDs are
affected by many factors such as the crystalline quality of
semiconductor device, imperfections of design and fabrication,
quenching circuit, and operational conditions. Therefore, actual performance of these APDs
is always compromised and optimized for different applications.

In the past decade efforts have been made to characterize and
further improve APD performance on the single-photon level at
1550\,nm~\cite{LZCL96,RGZG98,RWROT00,His00,SRSZRW01,Kar03,RGGSWZ04,Ramirez06,Pellegrini06,Itz07,Liu07,Jiang07,Jiang08}.
Recently, integration of the quenching electronics for InGaAs/InP
APDs to an ASIC~\cite{ASIC,TSGZR07} has been implemented. The
measured results on some key parameters of APDs demonstrate active
quenching ASICs can efficiently improve the noise-efficiency
performance, and it has been shown that these APDs can work in a
free-running mode~\cite{TSGZR07}. However, full
characterization of APDs with the ASIC is still necessary to better
understand the improvements they provided. In this paper, we fully
test 4 InGaAs/InP APDs at 1550\,nm with an active quenching ASIC
operating both in the free-running and gated modes, and compare the
improvements with conventional electronics.

\section{The setup and the principle of the ASIC}
\begin{figure}
\centering
  \includegraphics[width=0.45\textwidth]{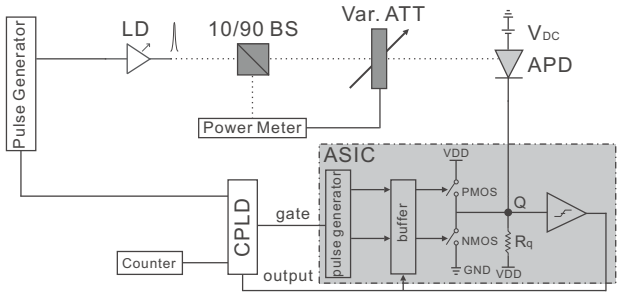}\\
  \caption{The experimental setup.}
  \label{setup}
\end{figure}

The schematic setup for testing APDs is shown in Fig.~\ref{setup}.
A digital delay pulse generator (DG\,535, Stanford Research Systems
Inc.) provides synchronous signals for the whole system. One of its
periodic outputs drives a 1550\,nm laser diode (LD) to produce short
optical pulses with $\sim 200$\,ps FWHM.
The optical pulses are split into two parts by a 10/90 asymmetric
fiber beamsplitter (BS). 90\% of the signal is monitored by a power
meter (IQ\,1100, EXFO Co.) to regulate the precise variable
attenuator (Var. ATT, IQ\,3100, EXFO Co.) in real-time and stabilize
the intensity of the output from the attenuator that goes to the
pigtailed APD. The pins of the APD and ASIC are soldered together on a
small printed circuit board, while the body of APD is fixed on
the top of 4-layer thermoelectric cooler and actively
stabilized with a closed-loop control.

The schematic diagram of the ASIC, fabricated with a $0.8\,\mu$m
complementary metal oxide semiconductor (CMOS) process, is shown in
the gray area of Fig.~\ref{setup}. The amplitude of the gate signals
from the complex programmable logic device (CPLD) is first converted
to the power supply voltage VDD ($+5$\,V) of the chip. Two pulses are then
generated to control the PMOS and NMOS switches respectively which
have extremely fast rise and fall times. There is a very short delay
between the two control pulses to avoid the simultaneous conduction
of the two switches. The timing is such that at the beginning of the
gate the PMOS switch is closed while the NMOS switch is opened, to
charge the voltage at the quenching point (Q) up to VDD, and then
the PMOS switch is reopened. The total voltage difference between
cathode and anode of the APD is $V_{bv}=$VDD$+|V_{DC}|$, exceeding
$V_{br}$ for Geiger mode operation. The NMOS switch remains open
until the end of the gate if no avalanche happens, or until the
active quenching after a triggered avalanche. During the avalanche
process, current across the APD rapidly increases and results in an
increasing voltage drop across the resistor $R_q$. The comparator
and the following circuit quickly detects the the voltage drop at Q
and immediately informs the buffer to close the NMOS switch to drop
the voltage at Q to zero, and also generate a synchronous detection
output to the CPLD. Normally the detector output maintains the high
level until the falling edge of the next gate. Actually, when a
detection is registered the CPLD inserts a short reset pulse after
the gate, otherwise the CPLD does nothing. In the free-running mode,
the gates from the CPLD are not used and VDD is applied to the
cathode of the APD until an avalanche is excited. Further technical
description about the ASIC can be found in Ref.~\cite{ASIC}.

\section{Performance tests of APDs}

We have tested 4 commercial APDs: \#1 (JDSU0131T1897);
\#2 (JDSU0122E1711); and \#3 (Epitaxx9951E9559) from JDSU; as well
as \#4 (PLI-DOI61910-040W059-076) from Princeton Lightwave, Inc.,
and compared the different performance characterizations of
these APDs with the ASIC quenching system.

\subsection{Integrated versus conventional quenching electronics}

\begin{figure}
\centering
  \includegraphics[width=0.45\textwidth]{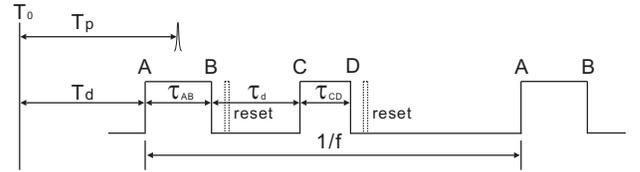}\\
  \caption{The timing diagram for the afterpulse measurements using the double-gate method.}
  \label{timing}
\end{figure}

\begin{figure}[h!]
\centering
a)
\includegraphics[width=0.45\textwidth]{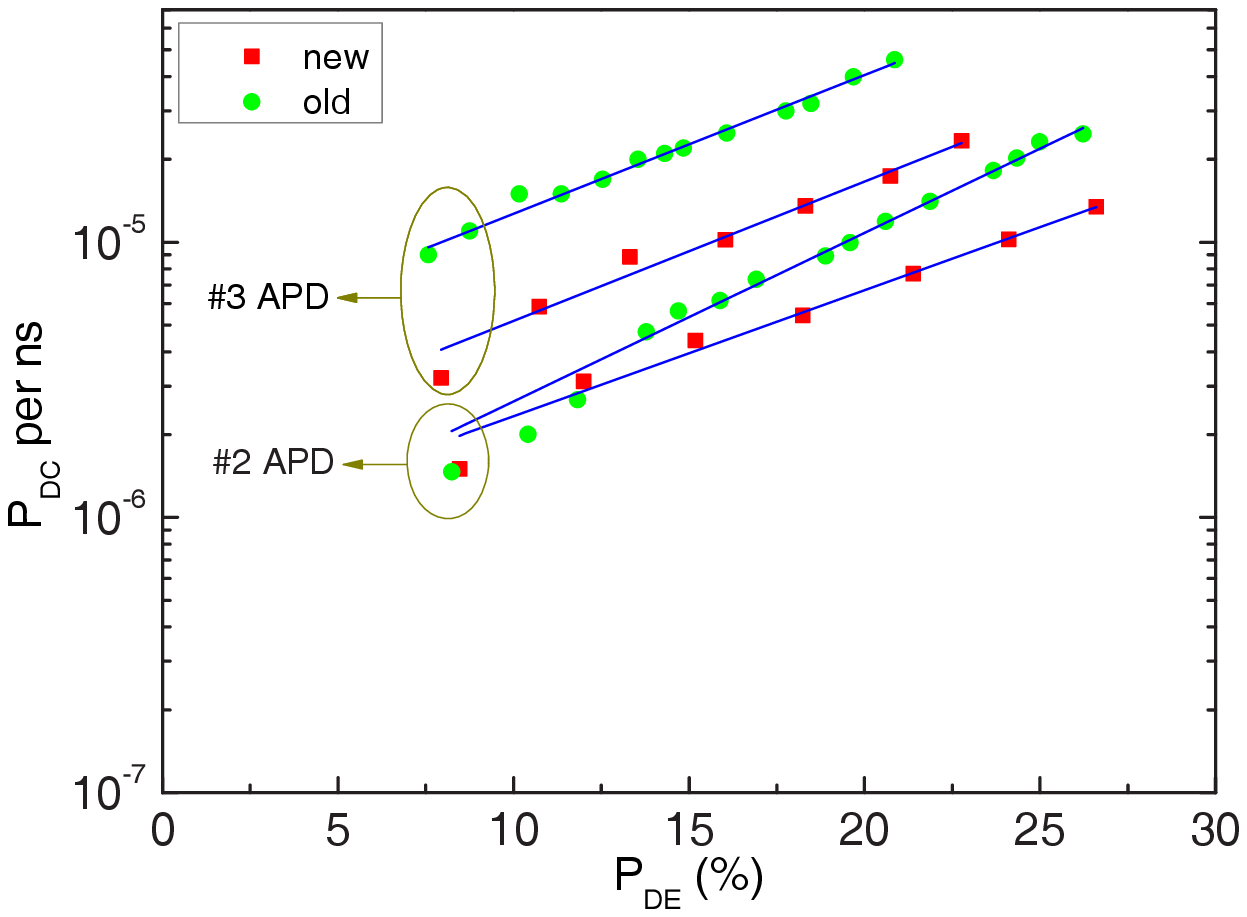}\\
b)
\includegraphics[width=0.45\textwidth]{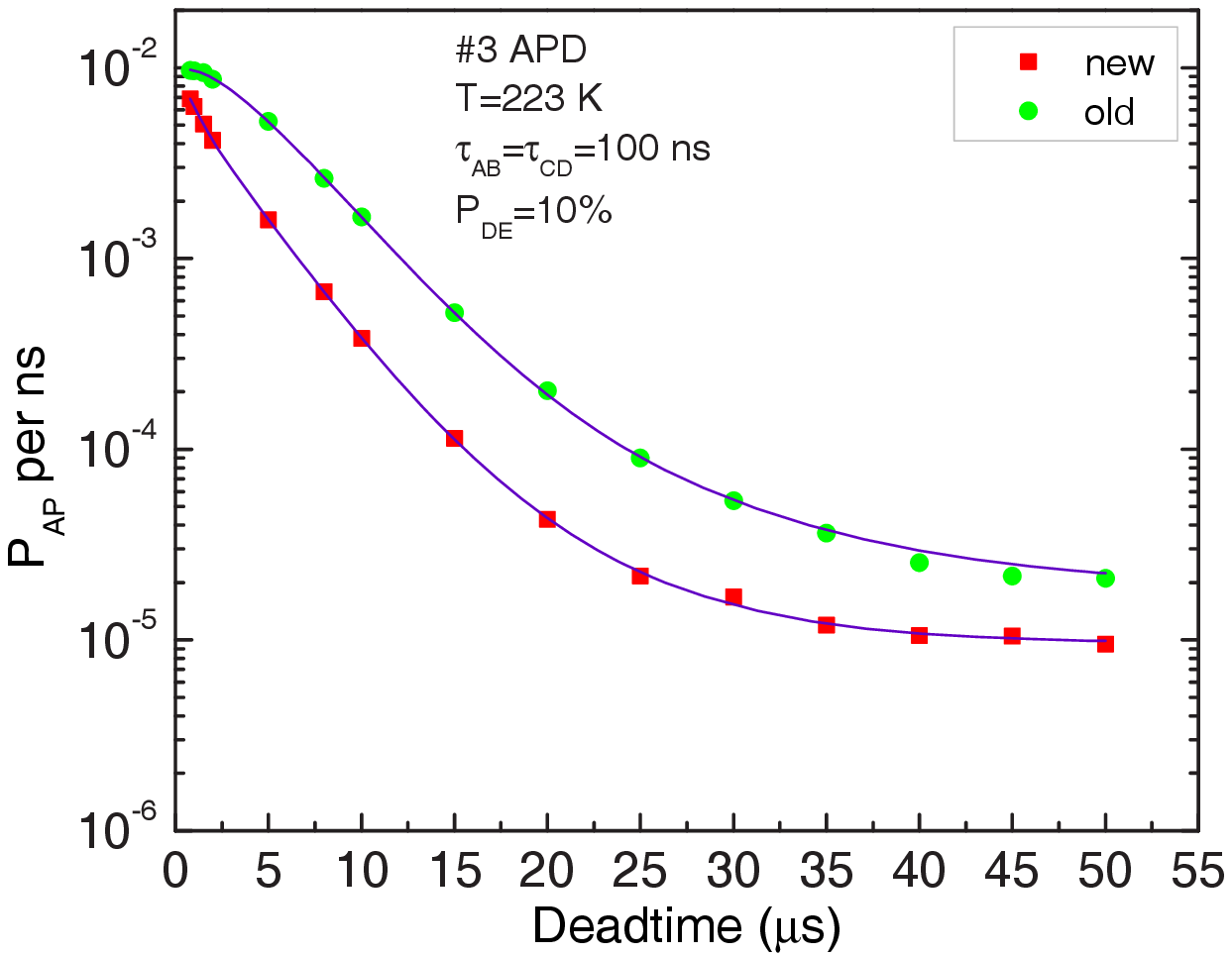}\\
c)
\includegraphics[width=0.45\textwidth]{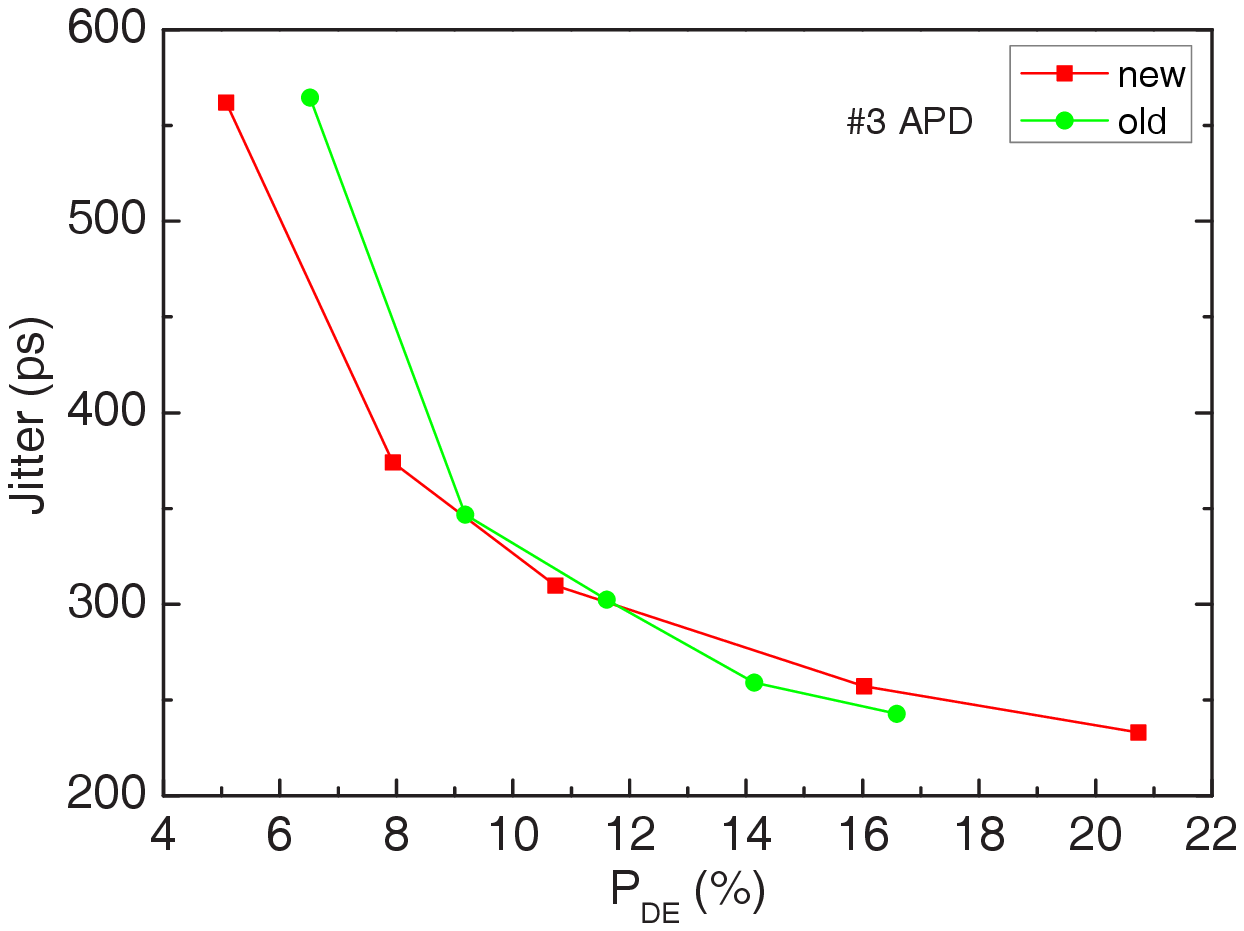}
\caption{a) Dark counts per ns ($P_{DC}$) versus detection efficiency ($P_{DE}$).
b) Afterpulse probability per ns ($P_{AP}$) versus deadtime ($\tau_{d}$.) c) Time jitter versus $P_{DE}$.}
\label{comp}
\end{figure}

Firstly, we perform the key parameter measurements on the same (\#3)
APD using the new (ASIC) and old (conventional non-integrated
circuit)~\cite{RGZG98,SRSZRW01} quenching electronics under the same
settings (T$=223$\,K ). Fig.~\ref{comp} shows the comparison results
for dark count ($P_{DC}$ per ns) vs single-photon detection
efficiency ($P_{DE}$) probabilities, afterpulse probability
($P_{AP}$) and jitter, respectively. Using the double-gate
method~\cite{SRSZRW01} (we discuss this in the latter section) as
shown in Fig.~\ref{timing}, these parameters can be related to:
\begin{equation}
\label{pa}
P_{DC}=\frac{C_{DC}}{f\tau_{AB}}, \,\,P_{AP}=\frac{C_{AP}}{C_{DE}\tau_{CD}}, \,\,P_{DE}=\frac{1}{\mu}\ln\frac{1-\frac{C_{DC}}{f}}{1-\frac{C_{DE}}{f}},
\end{equation}
where $C_{DC}$ ($C_{AP}$, $C_{DE}$) is the observed dark count
(afterpulse, detection) rate, $\tau_{AB}$ ($\tau_{CD}$) is the
effective width of detection (afterpulse) gate in ns and $\mu$ is the
mean photon number per optical pulse with repetition frequency of $f$.
During the experiment, the conditions are $f=10$\,kHz, $\tau_{AB}=\tau_{CD}=100$\,ns and $\mu=1$,
and these are fixed unless specifically mentioned in this paper.

The three curves manifestly exhibit the performance improvements
provided by the new quenching electronics. The improvement of a
factor of 3 in the $P_{DC}$-$P_{DE}$ performance for \#3 APD shown
in Fig.~\ref{comp}a is better than expected. As we know, due to the
ASIC the size of the electronics are greatly decreased and the
electronic cables and the lengths of wires are reduced. This brings
a lot of benefits such as superior signal integrity, minimized
parasitic capacitance and reducing fake avalanche signals due to
signal reflections or electronic noise. We also observe
$P_{DC}$-$P_{DE}$ performance improvements on other APDs, for
instance, for \#2 APD shown in Fig.~\ref{comp}a the ratio is always
about 1 (no improvement) when $P_{DE} < 13\%$ and slowly increases
to about 2 when $P_{DE} \sim 25\%$. The $P_{DC}$-$P_{DE}$
performance improvement ratio strongly depends on the APD devices
and operational conditions. Although the reasons of the significant
improvement for \#3 APD are not clear yet, one possibility could be
different gate heights and discrimination approaches between the two
quenching systems, as it is the noise that is improved here, for a
given excess bias voltage.

We see, in Fig.~\ref{comp}b, a significant improvement in the
$P_{AP}$ between the two cases as expected. The $P_{AP}$ is
generally proportional to the total number of carriers generated
during an avalanche and hence motivates small and rapidly quenched
avalanches. The results here clearly illustrate the circuit response
and quenching time of the new system for the avalanche
discrimination are faster than the old system. We will come back to
this in more details in the following sections.

Timing jitter (time resolution) is another key parameter.  It is
defined as the temporal uncertainty of detection output for an
avalanche with fixed arrival time of photons. Time jitter strongly
depends on device fabrication and $P_{DE}$, corresponding to excess
bias ($V_{eb}$) on the APD. Larger $V_{eb}$ can generate higher
electric fields, which will shorten the trapping time of the
carriers in the absorption and grading layers, and also the buildup
time of avalanche, hence reducing the jitter. To measure this we use
a time-correlated single photon counting (TCSPC) board (SPC-130,
Becker \& Hickl GmbH) with a time resolution of $6$\,ps FWHM and
minimum time slot of $815$\,fs, to measure the jitter properties. A
synchronized signal from a pulse generator is used as the TCSPC's
``stop'' while the detection output signal is used as ``start''. The
measured jitter is the overall jitter of the system, including the
jitter ($< 60$\,ps) and width ($\sim 200$~ps) of arrived optical
pulses, the APD's intrinsic jitter owing to the stochastic process
of carrier dynamics, as well as from the associated electronics. The
jitter performance is shown in Fig.~\ref{comp}c and we only see a
minor improvement when $P_{DE} < 10~\%$. We expect the electronic
jitter to be slightly better as the ASIC can efficiently reduce the
propagation time and jitter of the signals. At higher $P_{DE}$ we
don't observe the improvement and the negligible difference between
the two cases is due to contributions from the associated external
electronics, e.g., CPLD and discriminator that are used with
the new system but not the old one. However, varying degrees of
improvement have been observed on other APDs even at higher
$P_{DE}$.

\subsection{$P_{DC}$, $P_{DE}$ and thermal activation energy}
\begin{figure}
\centering
\includegraphics[width=0.45\textwidth]{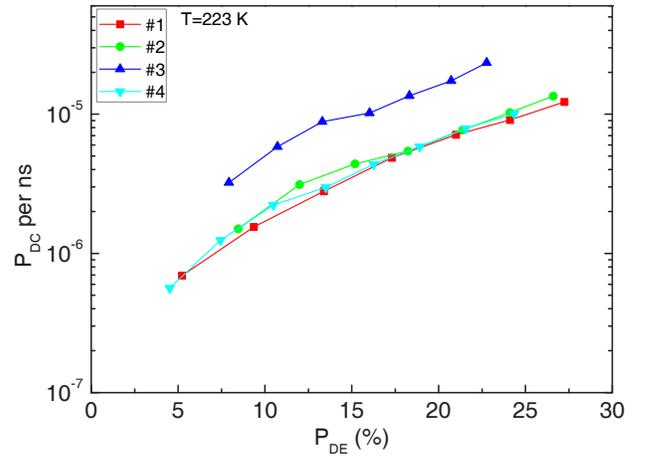}\\
\caption{$P_{DC}$ versus $P_{DE}$ of 4 APDs.}
\label{124}
\end{figure}

\begin{figure}
\centering
\includegraphics[width=0.45\textwidth]{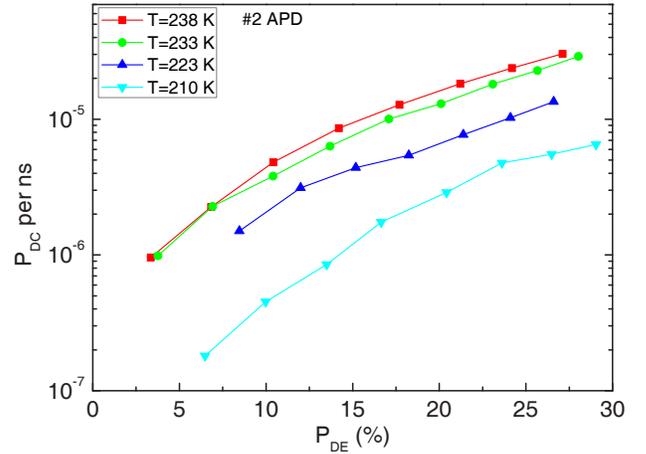}\\
\caption{$P_{DC}$ versus $P_{DE}$ of \#2 APD at different T.}
\label{2}
\end{figure}

In order to illustrate the universal improvements afforded by this
new quenching system, we use the new system operating in the gated
mode to repeat the measurement on different APDs and temperature
settings, as shown in Fig.~\ref{124} and Fig.~\ref{2}.

The $P_{DC}$-$P_{DE}$ behavior of \#1, \#2 and \#4 APDs are very
similar, with $P_{DC} \sim 1.6\times 10^{-6}$\,ns$^{-1}$ and
$P_{DE}=10\%$ at $223$\,K, as shown in Fig.~\ref{124}, but much
better than \#3 APD. Fig.~\ref{2} shows the $P_{DC}$
behavior of \#2 APD from $210$\,K to $238$\,K, and we see a
reduction in $P_{DC}$ to  $4.5\times 10^{-7}$\,ns$^{-1}$ for the same
$P_{DE}$.

\begin{figure}
\centering
\includegraphics[width=0.45\textwidth]{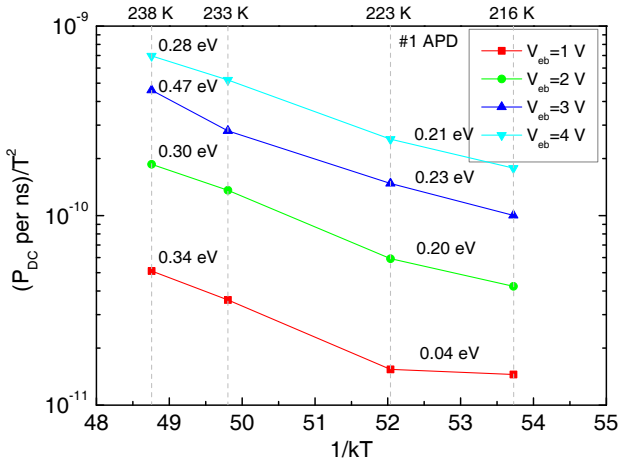}\\
\caption{Plot of $P_{DC}/T^{2}$ as a function of $1/k$T for \#1 APD.}
\label{act}
\end{figure}

The origin of the dark counts is mainly due to the defect
concentration in the semiconductor device. There are two main
mechanisms for the generation of dark carriers: thermal generation;
and tunneling generation. The thermal generation means that a
carrier is transferred from the valence band to the conduction band
either directly or via the midgap defects, owing to the thermal
excitation. Tunneling generation means that a carrier tunnels
between the two bands, or it is trapped by a defect first and then
tunnels to the conduction band, which is also called trap-assisted
tunneling (TAT)~\cite{Jiang07,Jiang08}. Combinations of the two
mechanisms are normally not taken into account. The simulations for
$1.06\,\mu$m InGaAsP/InP APDs performed by Donnelly \emph{et
al.}~\cite{Donnelly06} show that TAT in the multiplication layer
dominates the $P_{DC}$ at low temperature, while at high temperature
the two mechanisms compete with each other. Unfortunately, the dark
count model for $1550$\,nm InGaAs/InP APD is more complicated than
this, though one can investigate the so-called thermal activation
energy ($E_{a}$) to identify the dominant
mechanism~\cite{Itz07,Liu07,Jiang07}. Theoretically, the
relationship between $P_{DC}$, $E_a$ and temperature (T) can be
expressed as~\cite{Liu07}
\begin{equation}
\label{Ea}
    P_{DC}\propto T^2e^{-\frac{E_{a}(T)}{kT}},
\end{equation}
where $k$ is the Boltzmann constant and $E_{a}$(T) is a function of
temperature with slow variations.
In Fig.~\ref{act}, four curves of $log(P_{DC}/T^{2})$ versus
$1/k$T for \#1 APD with different $V_{eb}$ values are plotted.
We evaluate the difference of $E_a$ values for two small temperature
ranges ($216$\,K $\sim 223$\,K and $233$\,K $\sim 238$\,K).
The fitting values are displayed in Fig.~\ref{act}.
The results clearly show that generally higher
temperatures induce larger $E_{a}$ values and
suggest that the thermal generation mechanism around
$238$\,K dominates $P_{DC}$ while the TAT mechanism is more significant
around $216$\,K, see also ref.~\cite{Itzler}.

\subsection{Afterpulsing }

During an avalanche process, due to a photon detection, dark count
effects, or afterpulsing itself, a carrier can be trapped by a
defect in the multiplication layer. This carrier may excite
another avalanche - an afterpulse, during subsequent gates.
This process severely limits the APD performance for high frequency
operation due to the need to apply long, typically $\sim$
10\,$\mu$s, deadtimes where the APD is inactive. There are two
methods to measure the $P_{AP}$ behavior. The first approach
measures the total noise behavior as a function of $\tau_{d}$. When
$\tau_{d}$ is large enough, say, $100~\mu$s, the $P_{AP}$ is
negligible and the measured noise is primarily due to dark counts.
After subtracting $P_{DC}$, the quantity of noise left can be
attributed to afterpulsing. This method was used in
Ref~\cite{TSGZR07} but, while straightforward, generally
overestimates $P_{AP}$.

The other approach, the double-gate method~\cite{SRSZRW01}, as used
in our setup is illustrated in Fig.~\ref{timing}. If there is a
click during the detection (AB) gate, the CPLD will also generate an
afterpulse (CD) gate after AB's reset pulse with a delay of $\tau_d$
to the falling edge of the AB gate. This corresponds to the
deadtime. The CPLD also generates a reset pulse for the CD gate only
when an afterpulse detection is registered during this gate. This
method directly measures $P_{AP}$.

Assuming a Poisson distribution, $P_{AP}$ can be expressed as
\begin{equation}
\label{pap}
    P_{AP}=(1-e^{-R_{AP}(\tau_d)\eta_{av}\tau_{CD}})/\tau_{CD},
\end{equation}
where $R_{AP}(\tau_d)$ is the detrapping rate at time $\tau_d$ and
$\eta_{av}$ is the avalanche probability. We
use a multiple trapping model (multiple detrapping times) to
describe $R_{AP}(\tau_d)$~\cite{Itz07,Liu07},
\begin{eqnarray}
\label{papmulti}
    R_{AP}(\tau_d)=\sum_{i}\frac{N_{i}}{\Delta t_{i}}e^{-\tau_d/\Delta t_{i}},
\end{eqnarray}
where $N_{i}$ is the number of trapped carriers at the end of the
detection gate with a detrapping time constant of $\Delta t_{i}$.
There are  single trapping models that use a single detrapping time
constant $\Delta t$ but in many cases this does not correspond to
the measured results. The multiple trapping model effectively fits
the measured results but some physical questions remain,
e.g., why only 2 detrapping time parameters are needed for
modeling one APD while 3 parameters are required for another
\emph{etc}. In fact, quantitive description and modeling for
$P_{AP}$ behavior is still an intractable problem.
\begin{figure}[h!]
\centering
\includegraphics[width=0.45\textwidth]{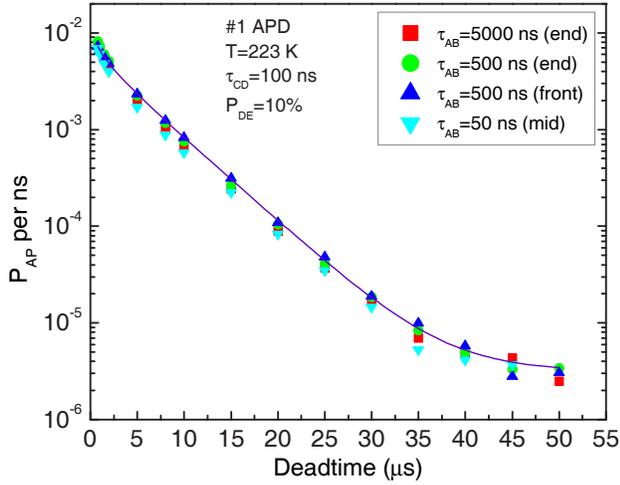}
\caption{$P_{AP}$ versus $\tau_{d}$.``end'', ``mid'' and ``front'' mean that
incident photons appear in the end, middle and front of AB gates, respectively.
The minimum $\tau_{d}$ is always $800$\,ns in Fig.~\ref{cd}-\ref{aptemp4}.}
\label{cd}
\end{figure}

To illustrate the suitability for free-running operations we look at
the $P_{AP}$ as we make our detection gates longer. The results for
\#1 APD are plotted in Fig.~\ref{cd} and fitted using the multiple
trapping model. $\tau_{CD}$ is fixed at $100$\,ns while $\tau_{AB}$
and the photon's arrival positions are altered. If the active
quenching was slow then the arrival position, or time, of the
photon's appearance in the $AB$ gate would be reflected in the
$P_{AP}$ behavior. A photon creating an avalanche at the start of a
long gate would generate more carriers, increasing the chances for
subsequent afterpulses, than in the short gate regime or if the
photon arrived at the end of a gate. The overlapping curves show
that the $P_{AP}$ behavior doesn't change for long gates, nor is it
dependent on the arrival time, and hence shouldn't change when we
move to a free-running regime.
\begin{figure}[h!]
\centering
\includegraphics[scale=0.7]{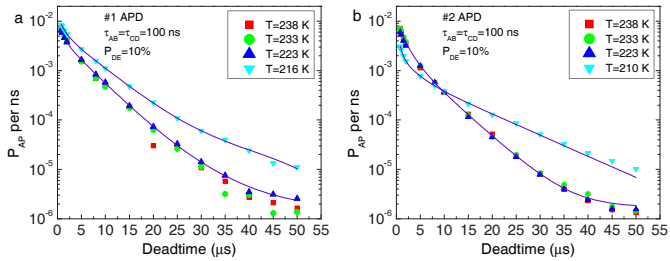}\\
\caption{$P_{AP}$ versus $\tau_{d}$ for \#1 (a) and \#2 (b) APDs at different T.}
\label{aptemp}
\end{figure}

\begin{table}[h!]
\center
\caption{\label{tab} The detrapping time parameters of fitting curves in Fig.~\ref{aptemp}.}
\begin{tabular}{|c|c|c|c|c|c|c|}
\hline
APD & T(K)      & $\Delta t_{1}$(ns) & $\Delta t_{2}$(ns) & $\Delta t_{3}$(ns)  \\   \hline  \hline
\#1 & 216       & 1135               & 5645               &                    \\   \hline
\#1 & 238-223   & 860                & 4385               &                    \\   \hline
\#2 & 210       & 615                & 2560               & 10135              \\   \hline
\#2 & 238-223   & 1020               & 2165               & 5075               \\
\hline
\end{tabular}
\end{table}

We finally study the temperature dependence of afterpulse.
The experimental results and fitting curves are shown in
Fig.~\ref{aptemp} and the fitting parameters are listed in
Table~\ref{tab}. When the temperature is varied from $238$\,K to
$223$\,K the $P_{AP}$ behavior is almost identical due to
the close trap lifetime parameters, but when the
temperature is at $216$\,K (\#1 APD) or $210$\,K (\#2 APD), there is
a distinct increase for the $P_{AP}$. The detrapping
lifetime can be modeled as~\cite{AP06}
\begin{equation}\label{dtr}
    \Delta t\propto e^{\frac{E_{ta}}{kT}}/{T^{2}},
\end{equation}
where $E_{ta}$ is the trapping activation energy. This formula means
that lower temperatures cause larger $\Delta t$ for traps,
corresponding to larger $P_{AP}$.

Moreover, when $\tau_d \lesssim 10\,\mu$s, the $P_{AP}$ of \#2 APD
at $210$\,K, in Fig.~\ref{aptemp}, is less than at other
temperatures, but the $P_{AP}$ of \#1 APD at $216$\,K is not.
According to the fitting results at $210$\,K, there is a trap type
with a fast detrapping lifetime of $615$\,ns in \#2 APD, which
causes rapid detrapping at small $\tau_d$, but when $\tau_d$ becomes
large, the effect of this trap type is gradually diminished while
the other trap types with $2560$\,ns and $10135$\,ns lifetimes start
to dominate the detrapping process. Unfortunately, this kind of fast
detrapping time is too short and/or too weak to be measured at the
other three temperatures and for \#1 APD.

\begin{figure}[h!]
\centering
  \includegraphics[width=0.45\textwidth]{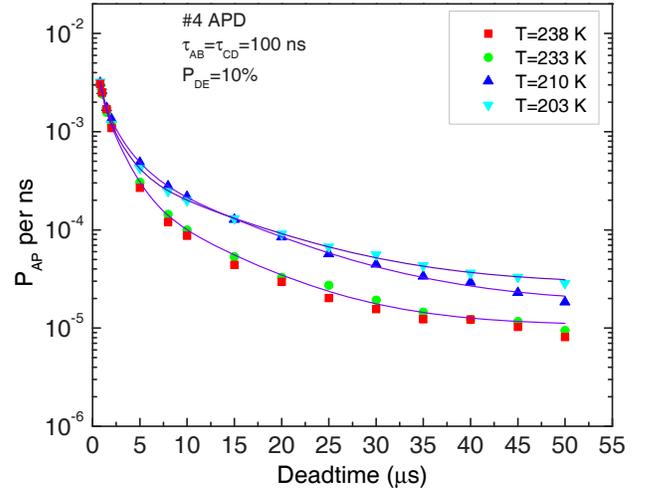}
   \caption{$P_{AP}$ versus $\tau_{d}$ for \#4 APD at different T.}
  \label{aptemp4}
\end{figure}

In order to validate the above phenomena, we perform the
measurements of $P_{AP}$ behaviors of \#4 APD from another
manufacturer, whose results are shown in Fig.~\ref{aptemp4}.
The $P_{AP}$ increases from $233$\,K to $210$\,K while the
cross point appears between $210$\,K to $203$\,K, which agrees well
with our explanation for the different $P_{AP}$ behaviors. We
believe that the $P_{AP}$ models so far are not perfect and further
investigations, including effective models and experiments, are
still needed.

\subsection{Free-running mode}
Free-running operation is very important for many applications such
as asynchronous and CW photon counting and quantum
cryptography~\cite{GRTZ02} \emph{etc}. Due to the lower noise
characteristics of InGaAs/InP APDs that use this active quenching
ASIC, some of us have recently been able to show that this is now also
possible for APDs in the telecom regime.
\begin{figure}[h!]
\centering
  \includegraphics[width=0.45\textwidth]{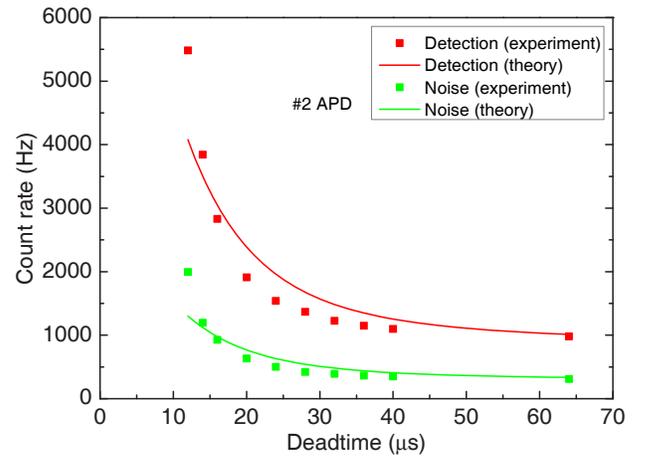}
   \caption{Plot of the detection and noise rates as a function of deadtime for \#2 APD at
   $V_{DC}=48.62$\,V, $N=10$\,KHz with CW photons and T$=210$\,K, operating in the free-running mode.}
  \label{free}
\end{figure}

As in the gated regime, the operation in the free-running mode
depends on the parameters of $V_{DC}$, $\tau_d$ and T.
However, unlike the gated mode, the afterpulse parameter in the free-running mode is more
difficult to evaluate.

As we said, with respect to Fig.~\ref{cd}, the $P_{AP}$ does not depend on the width of the gate,
which is applicable for the free-running mode. Indeed, it may not be obvious how the afterpulse probablity
evolves when the gate is open for such long times, though  it would appear that at worst,
the probability continues to decrease over the period of detection. Nonetheless,
we have previously seen that for short deadtimes the afterpulsing dominates \cite{TSGZR07}.
As we have now been able to use the double-gate method to characterize
the afterpulsing, in the gated regime, we can use a simple model to
describe the detection and noise rates for the free-running mode,
\begin{eqnarray}
\label{neteff}
    R=\eta N(1-\eta N\tau_{d})(1+\overline{P_{AP}}),
\end{eqnarray}
with $\eta=1-e^{-\mu P_{DE}}(1-P_{DC})$, considering the Poisson distribution. $P_{DE}$ and $P_{DC}$ are the detection
efficiency and dark count probability, and $N$ is the input photon
number. The term of $(1-\eta N\tau_{d})$ is for deadtime correction.
If we put $\mu$ = 0, we recover the noise rate.  $\overline{P_{AP}}$ is the total afterpulsing
contribution at $\tau_d$, calculated from integrating over the gated afterpulse
probability from $\tau_d$ to infinity (in practice 100\,$\mu$s is sufficient).
Fig.~\ref{free} shows the experimental rates as well as the results of our model as a function of $\tau_d$.

It is clear that a more complicated model is warranted. However,
the physics of these limitations is clear. In the small $\tau_d$ region we underestimate the rates
as we do not take account of cascaded afterpulses, i.e., higher order effects.
The more interesting region, from 20\,-\,40\,$\mu$s, we are overestimating due to the
difficulty in defining an appropriate integration range,
which will also change as a function of the photon flux, the intrinsic detection efficiency and the deadtime.
Importantly, we can also conclude that for small $\tau_d$,
if $N$ increases, then the $\overline{P_{AP}}$ value will decrease,
since photon clicks will increase while the multiple afterpulsing effects will be relatively less likely.

Our model makes a first attempt to both understand the afterpulsing and to develop a reliable
technique for determining the detector's characteristics, without resorting to complicated techniques
in a double-gate regime, there is still some way to go. Although the apparent need for large $\tau_d$ that,
in turn, limits the maximum count rate, this is highly dependent on the photon flux to be detected and
free-running APDs are certainly highly advantageous for applications with low to moderate count rates.

\subsection{Charge persistence}
Charge persistence is not normally a problem for synchronous
detectors  as the photons arrive during the gate. However, what happens
if a photon arrives before the gate is applied, as is possible in the
free-running mode, before the APD is activated after a deadtime?

When the detector is ``off'', i.e., at  $V_{bv}$ below
$V_{br}$, with only a few volts so that primary dark carriers can
still be generated and multiplied by the average dc gain but with a
small probability. When the gate pulse arrives some of the carriers
that remained in the multiplication layer can induce
avalanches~\cite{Kang03}. This is called ``charge persistence''
(CP), or sometimes referred as the ``twilight effect''~\cite{PM07}.
Similarly, when the CP carriers are released before the gate pulses
with the time difference less than the effective transit time, they
can also create afterpulses~\cite{Kang03}. Now let us consider
another case, where photons always appear before the gate. Based on
the same principle, in this case the number of dark CP carriers will
be increased and the CP effect will be expanded.

\begin{figure}[h!]
\centering
  \includegraphics[width=0.45\textwidth]{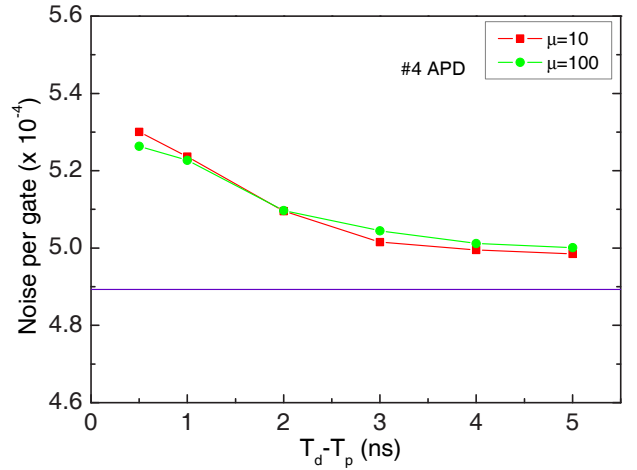}\\
   \caption{The noise, including CP and dark counts per gate, normalized by $\mu$, as a function of time difference
   between detection gate (T$_d$) and photons (T$_p$). The horizontal line is the dark count level.
   The results are tested using \#4 APD at $P_{DE}$=10\%.}
  \label{cp}
\end{figure}

We experimentally test this effect and the results are shown in
Fig.~\ref{cp}. By varying the time difference between the arrival
times of gates (T$_d$) and photons (T$_p$), we observe the changes
of the normalized (for $\mu$) noise per gate, for \#4 APD.
The two almost identical behaviors show that
the CP effect is proportional to photon numbers and,
per photon, can generate noise of about $10\%$ of the dark count level
with the time difference less than $1$\,ns.
When the time difference is larger than $\sim 5$\,ns, the CP effect
is negligible due to the characteristic exponential decrease.
Moreover, through using TCSPC, we also observe the detection events at the beginning
of the gate are more than those at other regions.
The CP effect will cause nonnegligible noise in
the case of high frequency gating or asynchronous high flux detection.

\subsection{Quenching time}

\begin{figure}[h!]
\centering
  \includegraphics[width=0.45\textwidth]{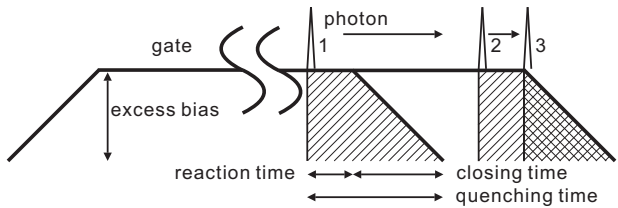}\\
   \caption{The principle of measuring the quenching time.}
   \label{QTscheme}
\end{figure}

Quantifying the quenching time, including the circuit reaction time and gate closing
time as shown in Fig.~\ref{QTscheme}, of an avalanche is very important to
understand the avalanche dynamics of APDs. Although an active
quenching ASIC should have a faster quenching time than conventional
electronics this has not previously been measured. More generally,
these results are also pertinent for rapid gating schemes that use
very short gates and hence terminate avalanches very quickly.

\begin{figure}[h!]
\centering
  \includegraphics[width=0.45\textwidth]{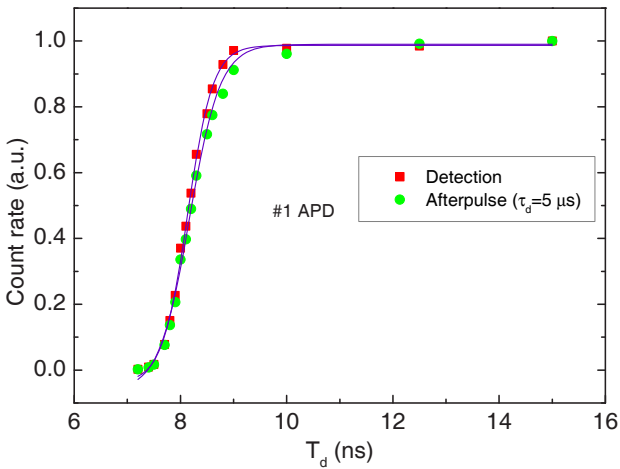}\\
   \caption{The count rates of detections and afterpulses with $\tau_d=$5\,$\mu$s versus the delay of detection gate (T$_d$).
   Points and lines are experimental values and theoretical S fits, respectively.
   The results are tested with \#1 APD at T$=223$\,K and $P_{DE}$=10\%.}
   \label{qt}
\end{figure}

The principle for measuring the quenching time is to compare
the count rate behaviors for detections and afterpulses, see Fig.~\ref{QTscheme},
using the double-gate measurement electronics. The total number of carriers,
during an avalanche, should be proportional to the excess bias on
APD and the excess bias duration, or the integral of excess bias over
the quenching time. Now we consider the case where photons arrive at the end of the detection gates, by
delaying photons. From phase 1 to phase 2 in Fig.~\ref{QTscheme},
the count rates of detection and afterpulse are both almost constant,
while from phase 2 to phase 3 the detection rate is still constant but
the afterpulse rate decreases first due to the decrease of the integral.
The time difference between the two phases can be regarded as the reaction time,
to detect the onset of the avalanche and send the signal to close the NMOS switch.
After phase 3, both of the rates drastically decrease until the end of
the closing time.

Fig.~\ref{qt} shows the results of these measurements on \#1 APD. From the slope
of the detection rate, we can obtain the closing time of the gate, which is only
around $1$\, ns. Although it is very hard to determine a precise value of the
reaction time from the fitting results, the slight shift between the detection
and afterpulse rates indicates that the reaction time is
much less than the closing time.

\section{Conclusion}
In summary, we have fully characterized an active quenching ASIC and
compared its operation with a conventional electronic circuit. To
show the improvements are universal we also characterized and
compared 4 different InGaAs/InP APDs. The APDs operating in the
gated mode exhibit substantial performance improvements compared
with the conventional quenching electronics and allow for
free-running operation. We also extract thermal activation energies
to identify the dominant mechanism of dark counts, and by employing
the multiple detrapping model in the gated mode and proposed model
in the free-running mode the afterpulse behaviors are well
illustrated. Moreover, we have characterized the charge persistence
and quenching time. The advantages of low afterpulsing and noise in
both theses regimes are mostly attributed to the state-of-the-art
ASIC.

\appendices
\section*{Acknowledgment}
The authors would like to thank Dr. A. Rochas and Dr. M. A. Itzler
for useful discussions.

\end{document}